\documentclass[acmsmall]{acmart}
\usepackage{libertine}
\usepackage{colortbl} 
\usepackage{placeins}
\usepackage[utf8]{inputenc}
\usepackage{amsmath}
\usepackage{pifont}

\DeclareUnicodeCharacter{03C7}{$\chi$}
\let\oldBbbk\Bbbk
\let\Bbbk\relax
\let\Bbbk\oldBbbk

\newcommand{\cmark}{\text{\ding{51}}}
\newcommand{\xmark}{\text{\ding{55}}}
\usepackage[many]{tcolorbox}    
\newtcolorbox{mybox}[1]{%
    tikznode boxed title,
    enhanced,
    arc=0mm,
    interior style={white},
    attach boxed title to top center= {yshift=-\tcboxedtitleheight/2},
    fonttitle=\bfseries,
    colbacktitle=white,coltitle=black,
    boxed title style={size=normal,colframe=white,boxrule=0pt},
    title={#1}}

\AtBeginDocument{%
  \providecommand\BibTeX{{%
    \normalfont B\kern-0.5em{\scshape i\kern-0.25em b}\kern-0.8em\TeX}}}

\setcopyright{acmlicensed}
\copyrightyear{2024}
\acmYear{2024}
\acmDOI{XXXXXXX.XXXXXXX}

\begin{document}

\title{Developers' Perceptions on the Impact of ChatGPT in Software Development: A Survey}


\author{Thiago S. Vaillant}
\email{thiagovaillant@hotmail.com}
\affiliation{%
  \institution{Federal University of Bahia}
  \city{Salvador}
  \state{Bahia}
  \country{Brazil}
}

\author{Felipe Deveza de Almeida}
\affiliation{%
  \institution{Federal University of Bahia}
  \city{Salvador}
  \country{Brazil}}
  \state{Bahia}
\email{fdeveza77@gmail.com}

\author{Paulo Anselmo M. S. Neto}
\affiliation{%
  \institution{Federal Rural University of Pernambuco}
  \city{Recife}
  \country{Brazil}}
  \state{Pernambuco}
\email{paulo.motant@ufrpe.br}

\author{Cuiyun Gao}
\affiliation{%
  \institution{Harbin Institute of Technology, Shenzhen}
  \country{China}}
  \state{Shenzhen}
\email{cuiyungao@outlook.com}

\author{Jan Bosch}
\affiliation{%
  \institution{Chalmers University of Technology}
  \country{Sweden}}
  \state{Gothenburg}
\email{jan@janbosch.com}

\author{Eduardo Santana de Almeida}
\affiliation{%
  \institution{Federal University of Bahia}
  \city{Salvador}
  \country{Brazil}}
  \state{Bahia}
\email{esa@rise.com.br}

\renewcommand{\shortauthors}{Vaillant and Deveza, et al.}

\begin{abstract}
As Large Language Models (LLMs), including ChatGPT and analogous systems, continue to advance, their robust natural language processing capabilities and diverse applications have garnered considerable attention. Nonetheless, despite the increasing acknowledgment of the convergence of Artificial Intelligence (AI) and Software Engineering (SE), there is a lack of studies involving the impact of this convergence on the practices and perceptions of software developers. Understanding how software developers perceive and engage with AI tools, such as ChatGPT, is essential for elucidating the impact and potential challenges of incorporating AI-driven tools in the software development process. In this paper, we conducted a survey with 207 software developers to understand the impact of ChatGPT on software quality, productivity, and job satisfaction. Furthermore, the study delves into developers' expectations regarding future adaptations of ChatGPT, concerns about potential job displacement, and perspectives on regulatory interventions. 

\end{abstract}



\received{}
\received[revised]{}
\received[accepted]{}

\maketitle

\section{Introduction}\label{sec:intro}

In recent years, Natural Language Processing (NLP) has seen remarkable progress, driven by advancements in Large Language Models (LLMs) and transformer-based architectures \cite{radford2019language, khurana2023natural,wolf2020transformers, min2023recent}. These innovations have promoted NLP to the forefront of Artificial Intelligence (AI) research, revolutionizing various domains such as healthcare, finance, and customer service \cite{kalyanathaya2019advances}.
One notable application stemming from these advancements is ChatGPT \cite{openai2022}, a derivative of GPT-3 \cite{wolf2020transformers, brown2020language,vaswani2017attention,radford2018improving, floridi2020gpt}, which focuses on facilitating natural language interaction with human users. The capability to comprehend and produce text in a coherent and contextual manner has paved the way for a multitude of applications, spanning from virtual assistant systems to medicine, automatic responses, and code generation \cite{10062688,dave2023chatgpt,deng2022benefits,tian2023chatgpt}.

Previous research has delved into the effectiveness of tools such as ChatGPT and Github Copilot through objective metrics, such as success rates, code generation quality, and defect identification, as well as their limitations and issues \cite{ChatgptCodeRefinement,bang2023multitask,sobania2023analysis,Sakib,chen2023chatgpt,liu2024your,yetistiren2022assessing,jaworski2023study,peng2023impact,wong2022exploring}. Nevertheless, few studies have investigated the impact of these tools on professional tasks and exploring developers' perceptions of their practical use in software development.

We conducted an online survey with 207 software developers to determine their opinions and beliefs regarding what factors contribute to the acceptance or rejection of ChatGPT among software developers. We focused on two main research questions, aiming to understand the factors influencing developers' adoption or rejection of ChatGPT usage in their activities and what they think about its popularization impact on the job market:

\textbf{RQ1: What factors contribute to the acceptance or rejection of ChatGPT among software developers?} \textit{Rationale:} With this research question, we aim to explore the factors that influence software developers' decisions to accept or reject ChatGPT. We seek to identify critical elements that shape their attitudes towards ChatGPT, including usability, reliability, satisfaction, and perceived usefulness.

\textbf{RQ2: What potential impact do developers believe ChatGPT can have on the software job market?} \textit{Rationale:} We aim to gain a clear understanding of developers' perspectives regarding the potential impact of ChatGPT on the software job market. Furthermore, our goal is to understand their expectations regarding how ChatGPT might affect job opportunities, job security, and the overall landscape of software development careers, considering the increasing prevalence of tools like ChatGPT.

Additionally, we opted to split the research questions into 4 sub-questions, presented in the Study Design (Section \ref{sec:methods}). This approach allows us to indirectly address the research questions through the sub-questions, while also contributing to the formulation of the survey questions.

In terms of results, our study revealed that:
\begin{enumerate}
    \item Developers perceive ChatGPT as enhancing productivity (73\% of the sample) and job satisfaction (76\% of the sample), attributing its use to reducing repetitive tasks and implementation time.
    \item ChatGPT excels in tasks like code understanding, suggesting, and commenting but performs poorly in security patterns and bug fixing.
    \item Participants anticipate career opportunities and challenges due to tools like ChatGPT, including higher entry barriers for beginners and concerns about job security.
    \item There is a favorable inclination towards regulation of tools like ChatGPT in software development (53\% of the sample), driven by concerns about social impact, job market dynamics, security, and copyright issues.
\end{enumerate}

In summary, we make the following contributions:
\begin{enumerate}
\item Our research provides insightful information about how developers perceive tools such as ChatGPT, providing a nuanced understanding of their perceived benefits and limitations in real-world software development scenarios.
\item By examining developers' attitudes towards regulation and market impacts, our research contributes to discussions surrounding the ethical and practical implications of integrating AI tools like ChatGPT into software development workflows.
\item The identification of associations between demographic profiles and perceptions of ChatGPT usage sheds light on potential differences in how different groups of developers perceive and use these tools, contributing to a deeper understanding of their user perspectives and preferences.
\end{enumerate}

Section \ref{sec:relatedwork} discusses related work involving software engineering and artificial intelligence. In Section \ref{sec:methods}, we describe the construction of the survey and the methods used for analysis. Section \ref{sec:population} contextualizes the survey population, while Section \ref{sec:results} presents the results. Section \ref{sec:discussion} discusses the findings and the implications of the work, while Section \ref{sec:limitations} describes the limitations and threats to validity. Finally, Section \ref{sec:conclusion} presents the conclusion and directions for future work.

\section{Related Work}\label{sec:relatedwork}

\subsection{AI and Machine Learning on Software Development}

Recent studies have explored the relationship between artificial intelligence (AI) and software engineering (SE), providing insights into applying SE principles to AI system development. One study surveyed developers from 26 countries, highlighting differences in development practices between Machine Learning (ML) and traditional software systems. It identified uncertainty and randomness inherent in ML system development, likening it to scientific programming \cite{Wan-Zhiyuan}.

An empirical investigation analyzed GitHub projects involving AI and ML applications, providing valuable insights into community development methods and traits compared to projects outside the ML domain \cite{Gonzalez}. Additionally, a case study examined Microsoft software teams developing AI-based applications, revealing major obstacles encountered and effective strategies for overcoming them \cite{Amershi}. Another study discussed bias in machine learning systems, addressing ethical concerns by identifying its underlying origins and proposing specific resolution strategies \cite{Chakraborty}. Furthermore, a literature survey pinpointed software engineering design patterns tailored for machine learning applications, suggesting ways to enhance their practical adoption within the community \cite{Washizaki}.

While these studies focus on optimizing AI system development through SE principles, our research takes a distinct approach by emphasizing AI tool integration within SE. Carleton et al. \cite{Carleton} also delve into the convergence between AI and SE, exploring the implications for both fields. The authors discuss applying SE principles to enhance AI software development and the value that AI tools can bring to software engineering practices. Additionally, they provide organizational guidelines for integrating AI into software development processes, while also addressing ethical considerations and the human role in utilizing AI systems.


Alongside the growth of Artificial Intelligence and Machine Learning models and tools, several software engineering researchers have been investigating strategies to incorporate these tools into the software development process. In \cite{Shafiq}, a literature review was conducted on using Machine Learning in the software development life cycle stages (Requirements, Design, Implementation, Quality Assurance and Maintenance), aiming to classify and correlate these stages with the employed ML techniques, types (Supervised, Unsupervised and Reinforcement learning), and tools. The authors observed a clear preference among researchers for applying ML in the analysis and quality assurance stages, representing more than half of the papers.

An example can be observed in \cite{Tsoukalas}, where a study was conducted to evaluate at least seven machine learning algorithms for classifying and identifying code smells and elements that may generate technical debt in software projects. After training the algorithms, satisfactory results were achieved in classifying classes with high technical debt based on predefined metrics. The authors also emphasized the potential for professionals to use these models and metrics as supplementary tools for identifying aspects of technical debt in their projects.

In \cite{Feldt}, a taxonomy is proposed to assist researchers and professionals in classifying and evaluating potential AI applications in software engineering. By creating this taxonomy, the authors hope to give software developers an initial guide for thinking through the benefits and drawbacks of incorporating AI tools into the software development process. A similar work was conducted in \cite{Rajesh}, where the authors analyzed the integration of AI activities in the software development process, considering both the waterfall and agile perspectives.
\subsection{Large Language Models (LLMs)}
Since the emergence of new Language Model-based systems (LLMs) such as ChatGPT, several researchers in SE have been investigated the applicability of this tool in software development activities. One of the aspects that has captured the attention of researchers is the potential to use language models for generating code based on natural language specifications. In \cite{Jain}, Jain et al. discuss LLMs, their potential to increase productivity, and their limitations regarding quality assurance in code generation.

Ronanki et al. \cite{Ronanki} explored the use of ChatGPT for requirements elicitation by defining six questions and asking ChatGPT to elicit requirements based on these questions. The elicited requirements from ChatGPT were then compared with those proposed by domain experts in requirements engineering to evaluate their quality. The requirements elicited by ChatGPT received higher evaluations than those proposed by domain experts. However, ChatGPT demonstrated poor performance in terms of feasibility. Overall, the study concluded that ChatGPT elicited highly abstract, indivisible, consistent, correct, and comprehensible requirements compared to domain experts. Thus, the authors argue that these results support ChatGPT's potential as a promising tool in requirements elicitation.

Another similar work was conducted in \cite{Mosel}, where it demonstrated that pre-trained transformers with software engineering context information can bring significant advantages in tasks that require some form of natural language processing, such as documentation or requirements elicitation.

In \cite{Sakib}, Sakib et al. evaluated the code generation and debugging capabilities of ChatGPT using a dataset of programming problems obtained from LeetCode, covering a wide range of categories and difficulty levels. The study's findings revealed a success rate of 71.8\% in ChatGPT's responses, with most correct answers being provided in the initial attempt. Additionally, the researchers investigated whether ChatGPT could enhance its incorrect responses or offer correct solutions when provided with error feedback from LeetCode. However, despite the high success rate on the first attempt, ChatGPT only improved its incorrect responses in 36\% of the cases. The authors highlighted that ChatGPT demonstrated higher proficiency in handling problems with well-structured methodologies and clear patterns but showed lower efficiency on problems involving decision-making with diverse scenarios and test cases.

To understand developers' perception and usage of Copilot, a study involving 24 participants was conducted to assess the use of Copilot in code generation \cite{Vaithilingam}. Despite the results not indicating an improvement in task completion time or success rate, developers continued to use it daily, claiming that Copilot provided them with a good starting point to implement functionality and reduced the online search effort. 
Nevertheless, given their potential for addressing development tasks and computing problems, it is necessary to be aware of the quality of the responses, as these models will also replicate patterns, defects, and biases existing in the training data \cite{Ozkaya}.

Diverging from their methodology, our study conducted a qualitative analysis, exploring the integration of ChatGPT in software development. We emphasized developers' perspectives on its influence on productivity and quality, particularly in development tasks such as code generation. Additionally, we investigated its effects on job security and satisfaction, gathering insights from developers with varying levels of expertise and educational backgrounds.

Based on our study, participants generally perceive the integration of ChatGPT into software development positively, attributing it to enhanced productivity and job satisfaction. This is consistent with Jain et al.'s discussions \cite{Jain}, which underscore the potential of Language Model-based systems (LLMs) like ChatGPT to boost productivity. However, different from Ronanki et al.'s study \cite{Ronanki}, which demonstrated ChatGPT's superior performance in requirements elicitation compared to domain experts, our study did not directly assess ChatGPT's performance in this specific activity, as our focus was primarily on tasks related to code. Sakib et al. \cite{Sakib} reported ChatGPT's relatively high success rate in code generation; nonetheless, our findings suggest a more nuanced perspective. While participants acknowledged ChatGPT's potential, they emphasized the importance of caution, particularly in ensuring the accuracy and reliability of the generated code. This concern aligns with Vaithilingam et al.'s caution \cite{Vaithilingam} regarding Copilot's responses, underscoring the importance of efficient code review despite such tools' assistance.

In comparison to Stack Overflow's study \cite{stackoverflowsurvey}, which explored developers' sentiments on AI/ML tools, both studies show a favorable attitude towards integrating AI tools into development workflows. While Stack Overflow found that 70\% of developers are either using or planning to use AI tools, with 77\% expressing a favorable view, our study similarly found positive perceptions of ChatGPT among participants, emphasizing enhanced productivity. However, unlike Stack Overflow's findings, which indicated that developers with over 21+ years of experience are less likely to use or plan on using AI tools, our study did not find a statistically significant association between years of experience and the intention to incorporate ChatGPT.

While our study focused on ChatGPT, Liang's study \cite{Liang_gptsurvey} was focused towards exploring the challenges and successes found by developers in the use of AI programming assistants. Similar to our study, their findings indicate that some of the reasons for using AI programming assistants include the tools' ability to help developers reduce keystrokes and complete programming tasks quickly. Performing repetitive tasks and providing learning assistance were also cited as successful cases.

Their work also reports that one of the most significant reasons why developers do not use these tools is because it is too difficult to control them to generate the desired output. One of their key findings mentioned that participants were most concerned about these tools potentially infringing on intellectual property, a topic that was also recurrent among the participants of our study.

In addition to the points mentioned, other aspects across the studies converged to similar results, indicating that ChatGPT may share much in common with AI programming assistants like GitHub Copilot from the users' perspective.

\section{Study Design and Methods}\label{sec:methods}
To address our study objective, we started with the defined research questions and broke down each of them into 4 sub-research questions.  The subquestions and its rationale are:

\begin{enumerate}
    \item RQ1: What factors contribute to the acceptance or rejection of ChatGPT among software developers?
    \begin{itemize}
        \item RQ1.1: How do developers perceive the effect of ChatGPT on their productivity while working on software development? \textit{Rationale:
        } Examining developers' perceptions of how ChatGPT impacts their productivity during software development tasks, this sub-research question seeks to determine whether ChatGPT enhances their productivity in different tasks and if it helps mitigate challenges in software development productivity.
        \item RQ1.2: How do developers feel that their satisfaction is impacted by ChatGPT? \textit{Rationale:} Focusing on exploring how developers' job satisfaction is affected by their usage of ChatGPT, this sub-research question aims to investigate the impact of ChatGPT on both social and technical factors related to job satisfaction.
        \item RQ1.3: How does ChatGPT influence the time spent on specific software development tasks? \textit{Rationale:} This sub-research question delves into how developers perceive the impact ChatGPT can have on the time required to complete certain tasks related to software development. Additionally, it explores how this perception relates to their views on productivity and job satisfaction.
        \item RQ1.4: How do developers assess the quality of ChatGPT in the context of software development? \textit{Rationale:} By focusing on understanding how developers evaluate the quality of ChatGPT within the context of software development, this sub-research question aims to gain insights into its performance across different software development tasks and determine which aspects are most influential in shaping its acceptance or rejection.
    \end{itemize}
    
    \item RQ2: What potential impact do developers believe ChatGPT can have on the software job market?
    \begin{itemize}
        \item RQ2.1: How do developers perceive the potential impacts of ChatGPT on their profession? \textit{Rationale:} This sub-research question aims to understand developers' perceptions of how ChatGPT could affect their profession. We are interested in learning what developers think about the impact of tools like ChatGPT on their careers in the near future.
        \item RQ2.2: What do developers think about the regulation of AI systems like ChatGPT? \textit{Rationale:} Investigating developers' perspectives on the regulation of AI systems, particularly concentrating on ChatGPT, this sub-research question aims to comprehend their views on regulation and the reasoning behind their positions.
        \item RQ2.3: What do developers think about the influence of ChatGPT on potential job layoffs in the software market? \textit{Rationale:} This sub-research question explores developers' perspectives on the potential impact of ChatGPT on job layoffs within the software market. We aim to understand whether developers believe ChatGPT could contribute to job layoffs in the software market and the factors influencing their beliefs.
        \item RQ2.4: What do developers think about the relationship between ChatGPT and automation in software development? \textit{Rationale:} Focusing on understanding developers' perceptions of the relationship between ChatGPT and automation in software development, this sub-research question aims to explore their feelings about the increasing automation in software development facilitated by ChatGPT and related tools, as well as whether developers believe they could be replaced in some development tasks.
    \end{itemize}
\end{enumerate}

\subsection{Survey Construction}
We designed an online survey with developers to understand their perceptions and feelings about ChatGPT as an auxiliary tool in software development. The survey questions were also intended to collect and correlate developers' opinions, including questions that allowed for response segmentation into different perspectives, such as age, gender, professional experience, and education level, while preserving the confidentiality of the answers.

The survey's structure was carefully planned to optimize its dissemination across different online platforms to gather responses and insights from various sources, such as social networks, communities, forums, and email lists. By doing so, we aimed to make the survey easy and convenient to complete, with a duration of less than 20 minutes to attract a more significant number of responses \cite{kitchenham2008personal}.

Based on the defined research questions and the survey proposal, we designed the survey in five Sections. The first section provided an introduction with a brief description of the survey, its estimated completion time, acceptance terms, and ways to contact the authors.

The second section included questions aimed at collecting demographic information to segment the obtained responses based on the collected data. These questions encompass age, professional experience, gender, and education level.

The survey's core starts from section three, where we present questions related to how developers perceive the quality of ChatGPT in performing software development tasks, such as code generation and bug detection, as well as potential features for improvement that could be added in the tool.

Section four follows a similar proposal to the previous one but with questions addressing the factors that influence the adoption of ChatGPT as a work tool, considering aspects such as productivity and job satisfaction.

Section five concludes the survey by focusing on the possible impacts on the software industry, including questions about how developers perceive potential changes in the job market, ranging from the automation of some activities to possible mass layoffs.

To investigate how developers perceive ChatGPT and how they feel about it, we created a series of open-ended and closed-ended questions for the survey. Each question addresses how developers think about different approaches to ChatGPT, using the Likert scale for closed-ended questions and sometimes linking them to an open-ended question that serves as a justification for the response. When approaching topics that may have multiple interpretations (e.g., productivity), we provide the definition we will use before asking the question, seeking to ensure that the answers align with what was asked and not with the respondent's interpretation based on their experience.
We present a definition of productivity that relates to the cost or effort expended versus the value of the software produced \cite{mills1983software}. We did the same for other questions, presenting the definition of quality as the pursuit of excellence and ensuring efficiency of the product or process \cite{jones1991applied}, and job satisfaction as an affective and cognitive evaluation of the professional experience \cite{Brief}.
We used a list of challenges influencing productivity for the Likert scale questions, such as finding relevant information, legacy code, technical competency, insufficient training, etc. We also employed the following technical and social factors affecting job satisfaction for others Likert scale questions: repetitive work, perceived productivity, ability to achieve goals, time to complete tasks, learning useful skills, training for engineering technologies, impactful work, and stress \cite{Storey}.

All closed-ended survey questions are mandatory, while only one of the open-ended questions is mandatory. The idea was to ensure that all crucial questions were answered and that complementary questions did not become burdensome for the respondents.

Two pilot tests were conducted before the survey was published, resulting in several changes compared to the original study: eight questions were removed, seven were modified, and sixteen were added. These changes were motivated by the presence of questions that did not necessarily address what we desired, where we noticed questions that had more than one possible interpretation (e.g., questions related to the concept of productivity), and questions where the answers were purely based on speculation from the survey respondents. All the questions used in our survey are available in the paper's GitHub repository \footnote{https://github.com/gpt-impact/Paper-content}.

\subsection{Survey Distribution}
The dissemination and promotion of the survey were carried out in a decentralized manner by the research authors, with each one being responsible for promoting it in their respective country. The main strategy used involved reaching out to software developers within our own contact networks and requesting their participation in the research. Additionally, in order to obtain a larger number of responses, other forms of promotion were implemented, such as sharing the survey within developer communities. The communities were selected based on the number of active users and their relevance to the research objectives and topics of interest. These communities included groups focused on technical discussions, networking communities, and job posting and employment opportunity groups. In these environments, our approach involved identifying developers already engaged in the job market through their community posts and inviting them to participate in the survey and contribute to our research. For this purpose, platforms and social networks such as LinkedIn, Facebook, Twitter, Discord, and Reddit were used.
\subsection{Data Analysis}
\textbf{Statistical Approach} \textendash\ For data analysis, we used Pandas in a Python environment together with RStudio for statistical analysis. Some contingency tables were constructed between categorical variables such as demographic characteristics of the population and their respective opinions and beliefs. To check for statistically significant associations between the analyzed categorical variables, the Fisher exact test \cite{fisher} was chosen. The choice of the Fisher exact test was mainly due to the fact that some cells in the constructed contingency tables had frequencies less than 5, which traditionally makes the use of the $\chi^2$-test impractical \cite{ChiSquared}. To overcome the computational difficulty of calculating \textit{p}-values for contingency tables larger than $2 \times 2$, \textit{p}-value calculation was done using the Monte Carlo method approximation, with a number of iterations on the order of $1 \times 10^6$. To distinguish the relevance of the results, we considered only associations with $\textit{p} < 0.05$.
Data visualization was done using Matplotlib in Python to construct bar charts of the questions present in the survey and Likert tables. For visualizing associations, we employed the VCD package from the R library to generate mosaic plots, which allowed for better visualization of cells with discrepant Pearson residuals \cite{vcdpackage}.

\textbf{Open Coding} \textendash\ For the open-ended questions, an open coding approach was employed to analyze the responses provided by the participants. The responses to the 10 open questions were examined by two of the authors, with each independently evaluating the responses. The concepts used to label the data were either derived directly from the response content (in vivo codes) or abstracted by the researchers involved in the analysis \cite{saldana2021coding}. Through coding, responses could be segmented and grouped under categories, aiming to visualize emerging patterns and recurring themes within the data. Upon completion of the coding process, the authors discussed the codifications and the "Memos" assigned to each response, resolving discrepancies to generate a final coding scheme \cite{landis}.
\section{Population Contextualization}\label{sec:population}

Conducting a study focused on a specific area and achieving a completely randomized participant sampling can be challenging. However, to avoid sampling biases, we made efforts to attain diversity among the research participants. In this section, we outline how our population is distributed, with a focus on five aspects: gender, age, nationality, education level and professional experience.

The participants were spread out across 31 countries and 4 continents. The top three countries where the participants came from were Brazil with 107, United States with 22, and Germany with 14.
Table \ref{tab:demographics} provides an overview of how our participants are divided based on gender, age, professional experience and education level. At first glance, it becomes evident that 89.85\% of the participants are male. However, when dividing this group based on the number of years of experience as software developers, we highlight the diversity within this population. The largest group constitutes 26.34\% of the sample, while the smallest comprises 14.51\%.

In terms of age, our sample includes individuals ranging from 18 to over 60 years old, divided into five groups: the first group consists of those aged 18 to 25, the second group ranges from 26 to 40, the third spans from 41 to 60, the fourth is for those over 60, and the last group is for those who chose not to disclose their age. We observed the predominance of two major age groups, where participants aged 18 to 25 represented 42.02\% of the total sample, and participants aged 26 to 40 totaled 38.16\%. Together, they exceeded 80\% of the entire population.

\FloatBarrier
\begin{table}[!h]
\caption{Demographics of the study participants}
\label{tab:demographics}
\begin{tabular}{|lcccccc|}
\hline
\multicolumn{7}{|c|}{\textit{Demographics}} \\ \hline
\multicolumn{1}{|l|}{\textit{Population}} &
  \multicolumn{1}{c|}{\textit{Overall}} &
  \multicolumn{1}{c|}{\cellcolor[HTML]{FFFFFF}\textit{\begin{tabular}[c]{@{}c@{}}Less than 1 \\ year\end{tabular}}} &
  \multicolumn{1}{c|}{\cellcolor[HTML]{FFFFFF}\textit{\begin{tabular}[c]{@{}c@{}}1 to 2 \\ years\end{tabular}}} &
  \multicolumn{1}{c|}{\cellcolor[HTML]{FFFFFF}\textit{\begin{tabular}[c]{@{}c@{}}3 to 5\\  years\end{tabular}}} &
  \multicolumn{1}{c|}{\cellcolor[HTML]{FFFFFF}\textit{\begin{tabular}[c]{@{}c@{}}6 to 10 \\ years\end{tabular}}} &
  \cellcolor[HTML]{FFFFFF}\textit{\begin{tabular}[c]{@{}c@{}}More than 10 \\ years\end{tabular}} \\ \hline
\multicolumn{1}{|l|}{\textit{All responses}} &
  \multicolumn{1}{c|}{\textit{207}} &
  \multicolumn{1}{c|}{\textit{39}} &
  \multicolumn{1}{c|}{\textit{39}} &
  \multicolumn{1}{c|}{\textit{50}} &
  \multicolumn{1}{c|}{\textit{31}} &
  \textit{48} \\ \hline
\multicolumn{7}{|c|}{\textit{Gender}} \\ \hline
\multicolumn{1}{|l|}{\textit{Male}} &
  \multicolumn{1}{c|}{\textit{186}} &
  \multicolumn{1}{c|}{\textit{32}} &
  \multicolumn{1}{c|}{\textit{35}} &
  \multicolumn{1}{c|}{\textit{49}} &
  \multicolumn{1}{c|}{\textit{27}} &
  \textit{43} \\ \hline
\multicolumn{1}{|l|}{\textit{Female}} &
  \multicolumn{1}{c|}{\textit{17}} &
  \multicolumn{1}{c|}{\textit{5}} &
  \multicolumn{1}{c|}{\textit{4}} &
  \multicolumn{1}{c|}{\textit{1}} &
  \multicolumn{1}{c|}{\textit{3}} &
  \textit{4} \\ \hline
\multicolumn{1}{|l|}{\textit{Non-binary}} &
  \multicolumn{1}{c|}{\textit{0}} &
  \multicolumn{1}{c|}{\textit{0}} &
  \multicolumn{1}{c|}{\textit{0}} &
  \multicolumn{1}{c|}{\textit{0}} &
  \multicolumn{1}{c|}{\textit{0}} &
  \textit{0} \\ \hline
\multicolumn{1}{|l|}{\textit{Preferred not to answer}} &
  \multicolumn{1}{c|}{\textit{4}} &
  \multicolumn{1}{c|}{\textit{2}} &
  \multicolumn{1}{c|}{\textit{0}} &
  \multicolumn{1}{c|}{\textit{0}} &
  \multicolumn{1}{c|}{\textit{1}} &
  \textit{1} \\ \hline
\multicolumn{7}{|c|}{\textit{Age}} \\ \hline
\multicolumn{1}{|l|}{\textit{18-25 years old}} &
  \multicolumn{1}{c|}{\textit{87}} &
  \multicolumn{1}{c|}{\textit{24}} &
  \multicolumn{1}{c|}{\textit{30}} &
  \multicolumn{1}{c|}{\textit{28}} &
  \multicolumn{1}{c|}{\textit{4}} &
  \textit{1} \\ \hline
\multicolumn{1}{|l|}{\textit{26-40 years old}} &
  \multicolumn{1}{c|}{\textit{79}} &
  \multicolumn{1}{c|}{\textit{10}} &
  \multicolumn{1}{c|}{\textit{9}} &
  \multicolumn{1}{c|}{\textit{20}} &
  \multicolumn{1}{c|}{\textit{24}} &
  \textit{16} \\ \hline
\multicolumn{1}{|l|}{\textit{41-60 years old}} &
  \multicolumn{1}{c|}{\textit{32}} &
  \multicolumn{1}{c|}{\textit{3}} &
  \multicolumn{1}{c|}{\textit{0}} &
  \multicolumn{1}{c|}{\textit{0}} &
  \multicolumn{1}{c|}{\textit{3}} &
  \textit{26} \\ \hline
\multicolumn{1}{|l|}{\textit{Over 60 years old}} &
  \multicolumn{1}{c|}{\textit{4}} &
  \multicolumn{1}{c|}{\textit{0}} &
  \multicolumn{1}{c|}{\textit{0}} &
  \multicolumn{1}{c|}{\textit{0}} &
  \multicolumn{1}{c|}{\textit{0}} &
  \textit{4} \\ \hline
\multicolumn{1}{|l|}{\textit{Preferred not to disclose}} &
  \multicolumn{1}{c|}{\textit{5}} &
  \multicolumn{1}{c|}{\textit{2}} &
  \multicolumn{1}{c|}{\textit{0}} &
  \multicolumn{1}{c|}{\textit{2}} &
  \multicolumn{1}{c|}{\textit{0}} &
  \textit{1} \\ \hline
\multicolumn{7}{|c|}{\textit{Education level}} \\ \hline
\multicolumn{1}{|l|}{\textit{Graduated High School or GED}} &
  \multicolumn{1}{c|}{\textit{39}} &
  \multicolumn{1}{c|}{\textit{15}} &
  \multicolumn{1}{c|}{\textit{7}} &
  \multicolumn{1}{c|}{\textit{14}} &
  \multicolumn{1}{c|}{\textit{1}} &
  \textit{2} \\ \hline
\multicolumn{1}{|l|}{\textit{Trade/technical school}} &
  \multicolumn{1}{c|}{\textit{29}} &
  \multicolumn{1}{c|}{\textit{8}} &
  \multicolumn{1}{c|}{\textit{10}} &
  \multicolumn{1}{c|}{\textit{6}} &
  \multicolumn{1}{c|}{\textit{4}} &
  \textit{1} \\ \hline
\multicolumn{1}{|l|}{\textit{Bachelor's Degree}} &
  \multicolumn{1}{c|}{\textit{86}} &
  \multicolumn{1}{c|}{\textit{15}} &
  \multicolumn{1}{c|}{\textit{19}} &
  \multicolumn{1}{c|}{\textit{23}} &
  \multicolumn{1}{c|}{\textit{15}} &
  \textit{14} \\ \hline
\multicolumn{1}{|l|}{\textit{Advanced degree (Master's, Ph.D.)}} &
  \multicolumn{1}{c|}{\textit{53}} &
  \multicolumn{1}{c|}{\textit{1}} &
  \multicolumn{1}{c|}{\textit{3}} &
  \multicolumn{1}{c|}{\textit{7}} &
  \multicolumn{1}{c|}{\textit{11}} &
  \textit{31} \\ \hline
\end{tabular}
\end{table}
\FloatBarrier

However, similar to the gender distribution, participants vary significantly in terms of their experience in software development. In the group of participants aged up to 25, we had 27.58\% with less than one year of experience, 34.48\% with 1 to 2 years of experience, 32.18\% with 3 to 5 years of experience, and 5.74\% with six or more years of experience. Meanwhile, the group of participants aged 26 to 40 showed a predominance of individuals with more experience in software development. Participants with up to 2 years of experience totaled 24.05\%; meanwhile 25.31\% had 3 to 5 years of experience, 30.37\% had 6 to 10 years of experience, and finally, 20.25\% had more than 10 years of experience.

The participants in our research were grouped into four categories based on their education levels. Among them, 18.84\% completed high school as their highest educational level, 14\% obtained at least a technical diploma as their primary qualification, and 41.54\% graduated from various colleges or universities with undergraduate degrees. Finally, 25.60\% hold one or more postgraduate degrees, including master's and doctoral degrees.

Among these groups, the postgraduate group stood out, with a significant concentration of 58.49\% of its members having more than 10 years of experience in software development. Meanwhile, the bachelor's degree group was the most diverse group, showing notable diversity in terms of experience. In this group, 17.44\% had less than one year of experience, 22.09\% had 1 to 2 years, 26.74\% accumulated between 3 and 5 years, 17.44\% had between 6 and 10 years, and finally, participants with more than 10 years of experience totaled 16.27\% of the members.

Regarding professional expertise, we had a wide range of professionals participating in the research. The group with the highest number of representatives was the full-stack developers, totaling 44.44\% of the sample. The second-largest group was the back-end developers, constituting 21.73\% of the participants. Data scientists made up 11.59\% of the sample, while front-end developers represented 11.11\%. Embedded application developers accounted for 4.83\% of the participants, and mobile developers totaled 2.41\%. Those involved in testing and QA activities reached 1.93\%. Finally, game developers, desktop application developers, real-time systems developers, and data engineers each represented 0.48\% of the sample.

In essence, our study, despite facing challenges in achieving a completely randomized participant sampling, has diligently addressed potential biases by providing a comprehensive breakdown of our population distribution. While a significant majority are male, the distribution across age groups, experience levels, educational backgrounds, and professional expertise demonstrate the complexity and richness of our sample.

\section{Results}\label{sec:results}
This section presents the results for each research sub-question identified in Section \ref{sec:methods}.



\textbf{RQ1.1: How do developers perceive the effect of ChatGPT on their productivity while working on software development?}

To address this research question, we examined developers' views on how ChatGPT influences productivity in software development. Our analysis considered the following aspects: finding relevant information, legacy code, technical competency, training, interaction with people, unclear requirements, poorly defined goals, and external dependencies.

Regarding the overall impact on software development productivity, according to the definition presented in \cite{mills1983software}, we found that 50\% (102/207) of the participants responded that ChatGPT somewhat improves productivity in software development, and 23\% (47/207) responded that it improves it significantly. This result is consistent with a previous StackOverflow \cite{stackoverflowsurvey} survey, where 33\% of participants judged increased productivity as the most significant benefit when using AI tools in software development \cite{stackoverflowsurvey}.

\begin{figure}[htbp]
    \includegraphics[width=0.6\textwidth]{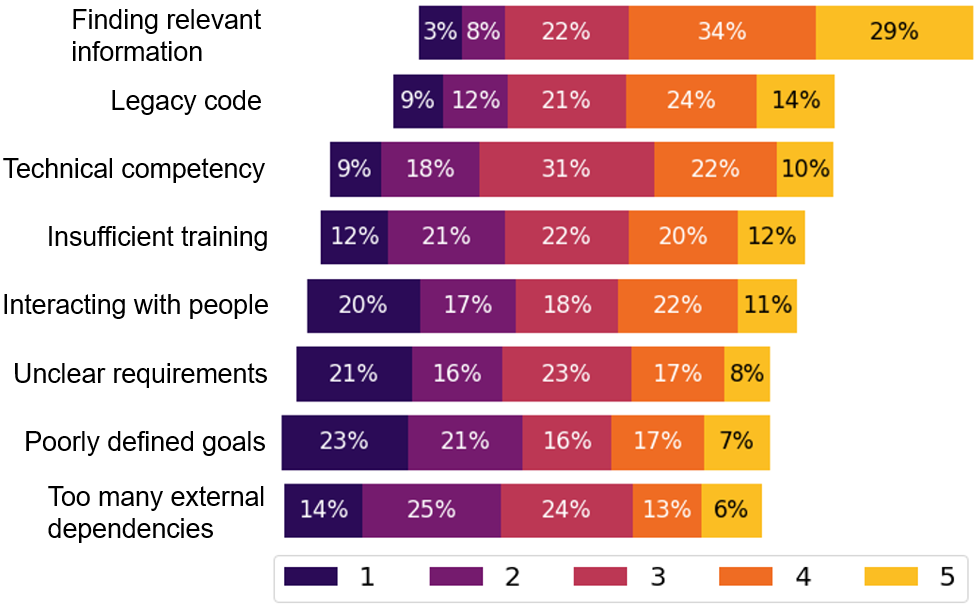}
    \caption{ChatGPT impact on developers productivity commom challenges}
    \label{fig:Productivity Likert}
\end{figure}

To gain deeper insights into this topic, we used a Likert-scale ranging from 1 (indicating minimal usefulness) to 5 (high usefulness). This approach aimed to assess ChatGPT's potential to mitigate challenges and obstacles hindering productivity in software development activities \cite{Storey}. Despite the majority of participants indicating that ChatGPT enhances their productivity, the average rating for this question was 3. This means, except for "Finding Relevant Information," which had an average rating of 3.8, developers do not consider ChatGPT to help mitigate the other issues. This suggests that the perception of increased productivity regarding ChatGPT may be linked to other factors (possibly subjective) within the software development activity. The distribution of responses for each of the challenges can be seen in Figure \ref{fig:Productivity Likert}.

Participants also provided feedback on the productivity impact in certain open-ended questions. Consistent with the overall findings, we received numerous responses highlighting increased productivity in software development. Based on the coding of the open-ended responses, other factors that developers may attribute to enhanced productivity with ChatGPT include "Lines of code," "Time," and "Features delivered."

\newcommand{\citatext}[1]{%
    \par
    \noindent\hspace*{1em}\begin{minipage}{\dimexpr\linewidth-2em}%
    \itshape #1%
    \end{minipage}%
    \par
}

In response to an open-ended question regarding the metrics utilized to evaluate the efficacy of integrating ChatGPT into their development workflow, the participant P136 explained how he became faster at coding and eventually saved time for other tasks with ChatGPT's assistance.:\\

\citatext{\cmark P136: "I would write code much slower without ChatGPT. Even though I modify its output heavily, I use it to generate a solid outline, help organize my thoughts, help with exception handling, and honestly integrate ChatGPT in every project I'm involved in."\\}

Participant P183 also reinforced it with a similar response:\\
\citatext{\cmark P183: "Features delivered faster (faster test cases, standard tasks are implemented faster in not contained in the framework) and higher code quality in the first place (bug avoidance)."\\}

This type of response was fairly prevalent and demonstrated a pattern, suggesting that speed and time allocation are potentially among the principal factors associated with the perceived increase in productivity facilitated by ChatGPT.

In the question regarding the motivations which participants integrate ChatGPT into their work, several responses highlighted the term "Repetitive Tasks." For instance, participant P32 remarked:\\
\citatext{\cmark P32: "Easy access to information if I ask the right questions. Automation of repetitive activities and simplicity."\\}

Furthermore, some participants pointed out that the level of productivity improvement largely depends on the type of task being carried out with ChatGPT. Overall, participants tend to believe that ChatGPT leads to a greater increase in productivity when used for simple and repetitive tasks, rather than complex and specialized tasks. For example, participant P128 highlighted:\\

\citatext{\cmark P128: "ChatGPT is excellent for some simple aspects of coding, but not great for complex codes that require a bit more knowledge. For basic tasks, it speeds up productivity, but for robust implementations with multiple stages, it can be challenging for it to understand what has been said previously."\\}

Participant P12 also commented something similar about the complexity of the tasks:\\

\citatext{\xmark P12: "My perspective varies depending on the complexity of the codes in question. For simpler codes, using ChatGPT can, in some cases, reduce the time needed to complete the task. On the other hand, for more advanced tasks, such as writing highly complex code, ChatGPT may indeed have limitations. It is not yet fully equipped to handle extremely complex challenges and may, in some cases, generate inadequate solutions. This means that, for highly complex programming tasks, it is more effective to rely on human experience and knowledge."}

\begin{mybox}{Finding 1}
Despite 50\% of the participants reporting that ChatGPT enhances their productivity to some degree, with 23\% indicating a notable improvement, we discovered that developers do not consider ChatGPT as helpful in addressing the challenges hindering their productivity. We highlight speed and time saved as key factors for the perceived productivity enhancement provided by the ChatGPT. In addition, ChatGPT is perceived to be more proficient in handling simple and repetitive tasks compared to complex and specialized programming tasks. The automation of repetitive tasks emerges as a notable factor motivating the integration of ChatGPT into their workflow. 
\end{mybox}


\textbf{RQ1.2: How do developers feel that their satisfaction is impacted by ChatGPT?}

We investigated the potential influence of ChatGPT on developers' satisfaction, in addition to assessing its impact on productivity. Our findings, aligning with the definition of job satisfaction as "An internal state that is expressed through the affective and/or cognitive evaluation of a job experience, with varying degrees of favor or disfavor" \cite{Brief}, indicate that 60\% (124/207) of the surveyed developers perceive ChatGPT to moderately enhance their satisfaction levels. Moreover, 16\% (33/207) expressed an opinion of an extremely positive impact, while an equivalent percentage 16\% (33/207) perceived no noticeable impact.

\begin{figure}[htbp]
    \includegraphics[width=0.6\textwidth]{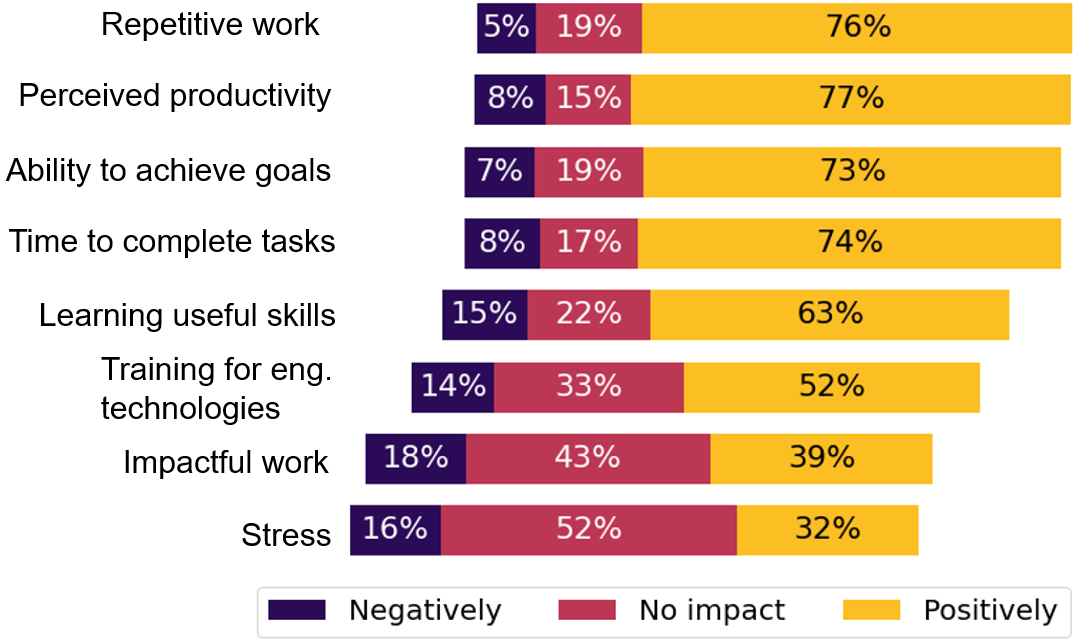}
    \caption{ChatGPT's influence on aspects of developers' job satisfaction}
    \label{fig:Satisfaction Likert}
\end{figure}

In addition, using a 3-point scale (Negative, No impact, and Positive) we investigated how ChatGPT influences the key technical and social factors that most significantly contribute to developers’ satisfaction (e.g., time to complete tasks) \cite{Storey}. In this inquiry, we listed 8 factors, all of which received more 'Positive' responses than 'Negative'. A particular emphasis was given to the factors "Perceived Productivity" and "Repetitive Work", with 77\% (160/207) and 76\% (157/207) respectively receiving positive responses. Interestingly, "Stress" was seen as the least affected factor by the participants, with 52\% (108/207) of responses indicating "No impact". Participants' responses to these and the other factors can be visualized in Figure \ref{fig:Satisfaction Likert}.

In the open-ended questions, several developers emphasized that ChatGPT significantly streamlines their work across various activities. Beyond the noted trend of reducing effort in repetitive tasks, the utilization of ChatGPT notably enhances the efficiency of the learning and research process. For instance, participant P147 
illustrates this by describing how he relied on ChatGPT to solve uncertainties in the absence of a more experienced developer to assist him: \\

\citatext{\cmark P147: "I'm a junior developer at a startup that had only one freelancer who built the application, and I was hired to make improvements and fixes on top of that. Due to the lack of technical support, as I'm the only developer in the team, I turned to tools that could assist me, acting as a senior, where I can ask questions and seek suggestions, but always needing to question because I make more mistakes than a senior... ChatGPT was, and still is, a great learning tool, especially when I started this job."\\}

In this context, a subset of participants remarked on the notably enhanced ease of conducting research with ChatGPT compared to conventional web search engines. Overall, participants stated that sending the question to ChatGPT is much faster in terms of cost-benefit to obtain an answer than navigating through multiple web pages seeking the desired answer. Participant P64 gave a response that illustrates this opinion well:\\

\citatext{\cmark P64: "Normally, I would waste a lot of time searching for something specific across countless websites, forums, documentation, and sometimes, not find the answer or guidance I need. With ChatGPT, not only do I get the answer, but I can also receive guidance, different viewpoints, explanations, insights, and even ask it to analyze as an expert or recommend learning materials. It's all positives. I'm considering purchasing access to the paid version soon, as it provides access to the most up-to-date database."\\}

\begin{mybox}{Finding 2}

60\% of participants perceived a moderately positive impact of ChatGPT on their satisfaction. Notably, it enhanced perceived productivity and reduced repetitive work, with stress being identified as the least affected factor. Furthermore, other responses highlighted ChatGPT’s facilitation of activities, support in learning, and provision of effective access to information compared to traditional tools.

\end{mybox}

\textbf{RQ1.3: How does ChatGPT influence the time spent on specific software development tasks?}

\begin{figure}[htbp]
\includegraphics[width=0.6\textwidth]{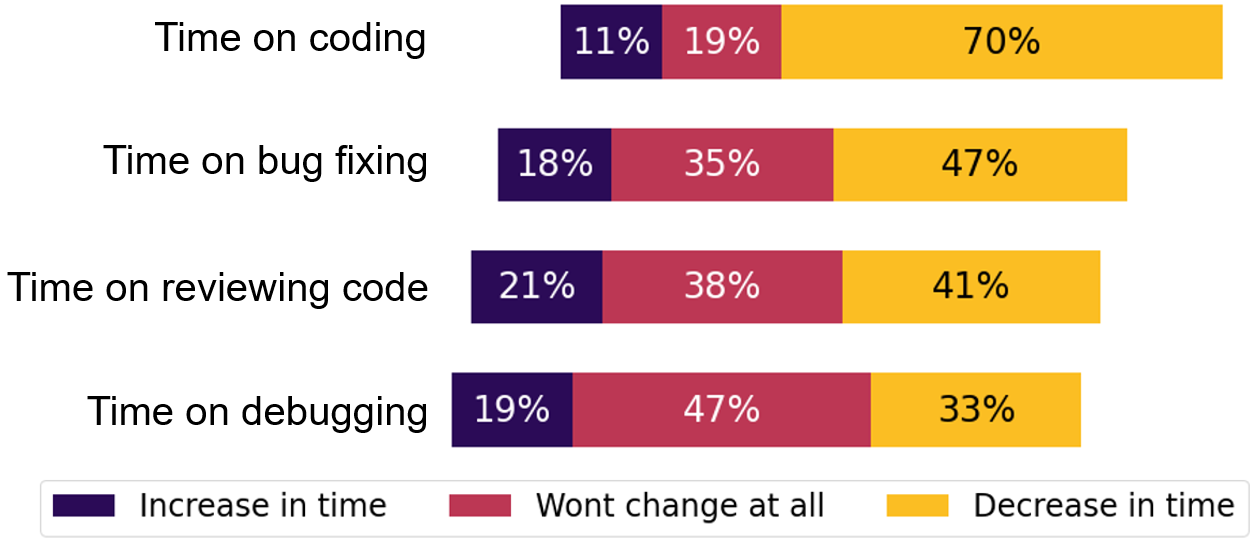}
    \caption{ChatGPT's Influence on Time Spent on Certain Task}
    \label{fig:Time Spent Likert}
\end{figure}

After examining productivity and satisfaction, we decided to investigate how ChatGPT influences the time required to complete a specific software development tasks, such as: coding, bug fixing, code review, and debugging. Our results indicate that out of the four programming activities listed, three presented a notable trend of time reduction when ChatGPT was used, particularly in coding. Figure \ref{fig:Time Spent Likert} shows participants' responses to these activities.

Some of the participants' open-ended responses provided valuable insights into the perception of ChatGPT's influence on activity completion time. For instance, the "Time" category emerged as one of the most frequently mentioned metrics when participants were asked about assessing the effectiveness of integrating ChatGPT into software development. Moreover, a notable proportion of participants highlighted a specific trade-off associated with using ChatGPT. While there is a significant reduction in coding time, the time required to review and fix the code generated by ChatGPT could equal or surpass the initial saved time. The following are some comments from some of the software developers that highlight these aspects:\\

\citatext{\xmark P50: "The only thing I see that might hinder a beginner in programming is the code review aspect. As mentioned above, coding, debugging, and fixing bugs may reduce time, but code review could increase it because sometimes people may not seek to understand what that code is doing or simply take the provided answer as is, which could compromise the project in the long run, perhaps with some features needing updates. That's why I think code review will increase the time."\\}

\citatext{\cmark P129: "I feel more productive because I spend less time actually coding, but I also spend more time reviewing. However, I would have to do the review regardless, even while coding, so I save that time."\\}

Some participants presented contrasting point of view, arguing that using ChatGPT in software development could potentially extend the time required to complete a task. They mentioned concerns such as hallucination, prompt engineering, and errors in the generated code as contributing factors to it. Participant P65 highlighted:\\

\citatext{\xmark P65: "I learned from my experience that explaining everything to ChatGPT and handling hallucinations takes way more time and distracts you from the focus. So, overall, I tend to spend way more time using ChatGPT for coding/debugging challenges than just staying focused and trying it myself (including Google searches, etc.)."\\}

Other participants remained skeptical about the possibility of increased productivity with the use of ChatGPT, considering that in certain scenarios, the use of ChatGPT could be detrimental. A comment from participant P80 illustrates this issue:\\

\citatext{\xmark P80: "I believe that sometimes more time is spent figuring out how to communicate with ChatGPT, testing the response, then, if there are issues, identifying the problem, and finally adapting it for the final objective. That said, sometimes it's simpler to use keywords for a possible solution or path on the internet. However, ChatGPT can assist by providing a more illustrated context when needed. The whole issue lies in the reliability of the response, where sometimes it's so well articulated that it deceives even if it's wrong. For these reasons, sometimes it's better to follow one's own line of reasoning rather than all these considerations."\\}

\begin{mybox}{Finding 3}
The findings reveal that a significant portion of participants reported that ChatGPT can reduce the completion time for specific programming tasks, notably on coding (70\%).
Participants stressed the importance of the ``Time'' category as a pivotal metric for evaluating the effectiveness of ChatGPT, recognizing a perceived trade-off between reduced coding time and the increased time needed for code review (21\% reported increased time for code review). While some reported heightened productivity, others cautioned that ChatGPT usage could potentially prolong task completion time due to challenges such as hallucination, prompt engineering, and code errors.

\end{mybox}

\textbf{RQ1.4: How do developers assess the quality of ChatGPT in the context of software development?}

Finally, still within the scope of RQ1, we aimed to understand how developers perceive the quality of ChatGPT within the software development context. As seen in some of the open responses presented in the previous RQs, the quality that developers attribute to ChatGPT strongly depends on the complexity level of the task they intend to perform with it. 

\begin{figure}[htbp]
\includegraphics[width=0.6\textwidth]{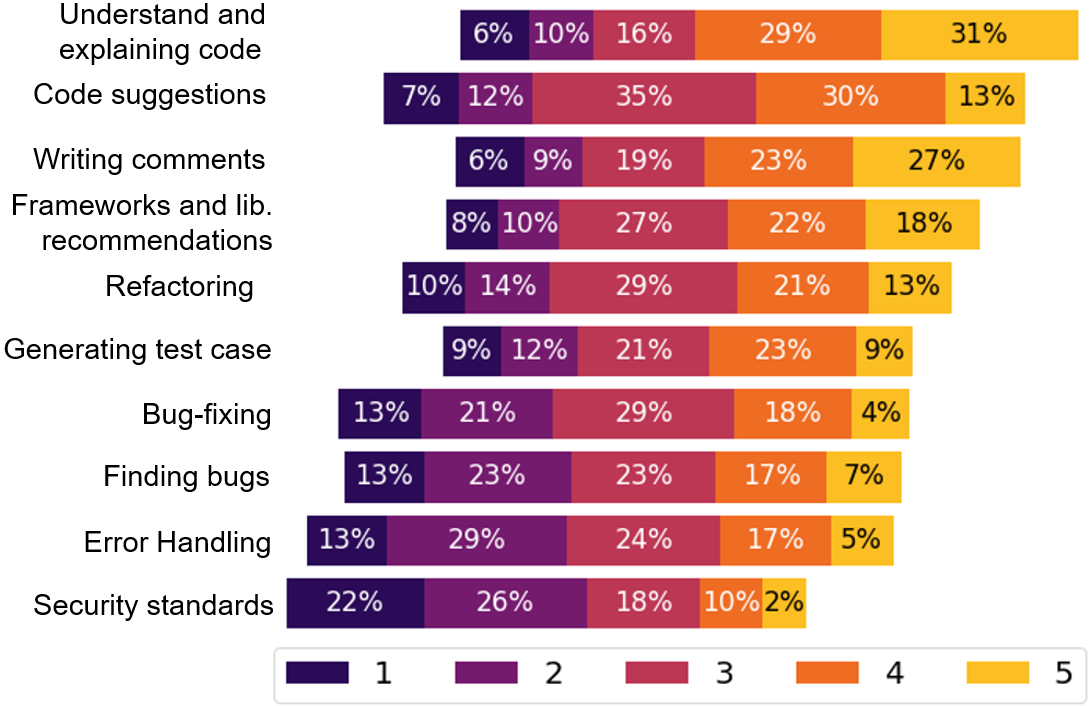}
    \caption{Quality of ChatGPT in tasks related to software development}
    \label{fig:Quality Likert}
\end{figure}

Thus, we asked the survey respondents about how they assess the quality of ChatGPT in various common software development tasks. The tasks most highly rated by participants, in order of preference, were: Understand and explaining code, Writing Comments, and frameworks and library recommendations. Meanwhile the worst rating belonged to security standards, error handling, and finding bugs. Figure \ref{fig:Quality Likert} shows the summary of the responses.  

We identified that according to survey respondents, ChatGPT had good results in specific tasks, notably in generating, explaining, commenting, and refactoring code—tasks primarily related to syntax and symbolic manipulation. On the other hand, tasks demanding a deeper comprehension of the code, such as error handling, bug fixing, and security patterns, received lower ratings (averaging below 3), which is in accordance with previous research on the low quality of ChatGPT in code review tasks \cite{ChatgptCodeRefinement}. This pattern is further reinforced in the open responses. When participants were asked about the challenges and limitations of ChatGPT, the most prevalent concerns included its outdated nature (Since the free version only contains data up to January 2022.), provision of incorrect information, tendencies toward hallucination, and potential security risks.

Based on the responses, a significant issue in working with newer libraries arises from the fact that the ChatGPT 3.5 Turbo language model, distributed for free by OpenAI, was trained on data only up until September 2021 \cite{openai2023}. Consequently, despite the "Code suggestions" functionality being positively rated, for any libraries or frameworks that have received updates since then, ChatGPT will inevitably be unable to provide accurate information or code, irrespective of the quality of the prompt. Participant P136 stated that:\\

\citatext{\xmark P136: "The largest issue so far is that sometimes ChatGPT has outdated information. For example, it is not up to date with the Twitter API and Tweepy. ChatGPT recommends old implementations of tweet scraping that are deprecated. Similar things happen with ASYNC PRAW, Django, and others. It works well on frameworks and libraries with minimal changes in the past 2 years."\\}

Hallucination is another problem that can make the developer's job even harder. Instead of simply stating it cannot fulfill a request, ChatGPT might create imaginary things and present them convincingly. This forces the user to carefully decide through what is true and what is not, which can consume a lot of the developer's time that could otherwise be spent on their actual tasks. The following are some comments from some of the software developers that highlight these: \\

\citatext{\xmark P135: "It is very easy to get ChatGPT to hallucinate the existence of entire libraries and programs. This makes distinguishing the correct parts from the incorrect especially hard."\\}

\citatext{\xmark P21: "An obvious limitation of the free level of ChatGPT is the lack of knowledge beyond 2021. This becomes problematic when requesting code snippets for languages like Rust or Go, as it often provides code with outdated content."\\}

Regarding security, discussions centered around two primary concerns: the possible security risks and weaknesses in the code produced by ChatGPT, and concerns about privacy breaches that could expose sensitive data. Participant P151 shared a situation where he noticed a simple security error in the code generated by ChatGPT:\\

\citatext{\cmark P151: "I feel good about the time reduction, but if done in moderation, it won't program for me, and attention is required because it makes silly security mistakes. As an example, I used it to create a user listing route. It returned all users from the database and then searched for the corresponding username and password on the front end, instead of making the mistake of sending the username and password as parameters to the backend."\\}

Other participants, such as P101, also expressed worries about data privacy, particularly considering proprietary code:\\

\citatext{\xmark P101: "Issues with code sharing. Every time it was necessary to remodel my problem to avoid sharing proprietary code on ChatGPT."\\}

\begin{mybox}{Finding 4}
Developers’ perceptions of ChatGPT's quality in software development vary significantly based on the complexity and nature of the task. They generally gave high ratings to tasks like code explanation, commenting, and refactoring, while tasks requiring a deeper comprehension of code, such as bug-fixing and adherence to security standards, received lower ratings. Participants also noted a limitation in ChatGPT's effectiveness stemming from outdated training data, resulting in inaccuracies, particularly with newer libraries and frameworks. Regarding quality, the primary challenges reported by developers include hallucination, security risks, and concerns about data privacy.
\end{mybox}


\textbf{RQ1. Association Analysis.} Initially, we observed that participants with advanced education degrees (M.Sc and Ph.D) demonstrated a higher inclination towards mistrusting ChatGPT, while those with lower educational levels (H.S and B.Sc) exhibited a greater inclination to fully trust ChatGPT ($p = 0.006$), indicating a potential association among the two variables. However, we found no statistically significant association between education level and the use of ChatGPT ($p=0.887$), nor with the intention to incorporate ChatGPT into software development ($p=0.359$).

Additionally, our analysis revealed that novice developers represent the group that uses ChatGPT most frequently ($p=0.033$). We also identified a strong association between higher Likert scores and the frequency of ChatGPT usage. Participants who reported using ChatGPT more frequently also rated positively the quality of \emph{Refactoring} ($p = 0.007$), \emph{Code comments} ($p < 0.001$), \emph{Error Handling} ($p < 0.001$), \emph{Code Understanding} ($p < 0.001$), and \emph{Code suggestions} ($p < 0.001$), as well as ChatGPT's ability to mitigate challenges related to \emph{Technical competency} ($p = 0.002$), \emph{Unclear requirements} ($p = 0.002$), \emph{Insufficient training} ($p = 0.002$), and \emph{Legacy code} ($p = 0.008$). In all instances, lower ratings were associated with participants who utilized ChatGPT infrequently or not at all.

Regarding trust, its absence is associated with low ratings of \emph{ChatGPT} in terms of quality, productivity challenges, and task time. Associations with higher statistical significance were found regarding the quality of \emph{Error handling} ($p < 0.001$), \emph{Code suggestions} ($p < 0.001$), \emph{Bug-fixing} ($p = 0.016$), \emph{Code comments} ($p < 0.001$), \emph{Understanding code} ($p = 0.003$), \emph{Refactoring} ($p = 0.006$), challenges of \emph{Technical competency} ($p < 0.001$), \emph{Legacy code} ($p=0.002$), \emph{Unclear requirements} ($p = 0.006$), and increased \emph{Time spent on reviewing code} ($p < 0.001$). More details about the analysis can be seen in the paper's GitHub repository.

\begin{mybox}{Finding 5}
Professionals with higher education levels maintain a certain skepticism regarding ChatGPT compared to others ($p = 0.006$). However, this lack of confidence does not seem to lower their inclination to use ChatGPT ($p=0.359$). The lack of confidence has been associated with low ratings on the quality of ChatGPT, especially in \emph{Error handling} ($p < 0.001$) and \emph{Code suggestions} ($p < 0.001$). On the other hand, excessive use of ChatGPT is more frequent among less experienced developers ($p=0.033$).
\end{mybox}



\textbf{RQ2.1: How do developers perceive the potential impacts of ChatGPT on their profession?}

To address this research question, we focused on two distinct aspects. Firstly, we explored developers' expectations regarding ChatGPT's impact on their careers in the upcoming years. Next, we asked if they would consider integrating ChatGPT into their work or daily routines.

We evaluated this topic using a combination of two closed-ended questions and one open-ended question. Regarding the career impact perspective, we used a closed-ended question for assessment while allowing participants who did not agree with any of the provided statements to express their opinions. The responses ranged from suggesting that ChatGPT would not impact their career to indicating that individuals who do not adapt to it could become obsolete, missing out on future job opportunities.

Among those who selected one of the provided alternatives, opinions were divided as follows: 41\% (86/207) of participants agreed that ChatGPT would require adaptation and the development of new skills in the coming years; 16\% (34/207) believed that ChatGPT would not have significant impacts on their careers in the foreseeable future; and finally, some anticipate that ChatGPT and similar technologies will create valuable opportunities for their careers in the next years totaling 38\% (79/207) of the answers.

\begin{figure}[htbp]
    \includegraphics[width=0.6\textwidth]{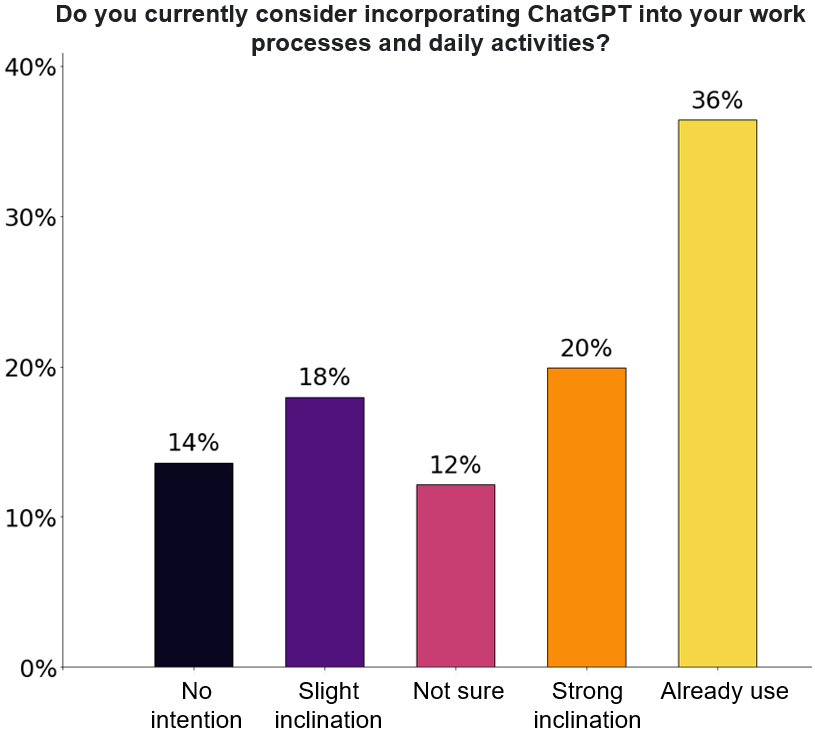}
    \caption{Views on Integrating ChatGPT into Work}
    \label{fig:Incorporating ChatGPT}
\end{figure}

Lastly, we asked if developers want to incorporate ChatGPT into their daily lives and work. As can be seen in Figure \ref{fig:Incorporating ChatGPT}, 36\% (75/207) reported that they already use ChatGPT in their work processes and daily activities, 20\% (41/207) confirmed they have a strong inclination to incorporate ChatGPT into their daily activities, 12\% (25/207) reported that they do not have a formed opinion on the matter, 18\% (38/207) said they have a slight inclination towards incorporating ChatGPT into their daily activities, and lastly, 14\% (28/207) of participants confirmed that they have no intention of incorporating ChatGPT into their daily lives or work activities.

Analyzing the arguments provided by the participants, there is considerable diversity in the reported usage patterns among those who claimed to use ChatGPT at work already. For instance, P7 predominantly uses ChatGPT in activities unrelated to programming.\\

\citatext{\cmark P7: "I already use ChatGPT in activities related to texts, emails, and documents where I see a real increase in productivity when using the tool, but I don't use it as much when I'm programming."\\}

Other participants explore the potential of ChatGPT in the realm of development, as exemplified by P36, who uses it to perform repetitive tasks and explore different implementation possibilities.\\

\citatext{\cmark P36: "I use it for repetitive tasks or to explore possibilities when implementing a new feature, and then I conduct further research"\\}

Meanwhile, P52 turns to ChatGPT to clarify doubts related to a programming language with which they are not familiar.\\

\citatext{\cmark P52: "I joined a company where I had never worked with the chosen language for new projects. It helped with the basic concepts and other doubts."\\}

Lastly, there are participants, such as P116, who utilize the generalist nature of ChatGPT, using the tool as assistance in a wide range of tasks.\\

\citatext{\cmark P116: "After understanding how ChatGPT works and starting to use it more frequently, I realized that it significantly helps me in various activities. Its ability to be a generalist and assist me in different types of tasks increases my confidence and comfort in using it more often. In the initial tests, when performing tasks I already knew how to solve, I was impressed. In addition to offering alternative solutions, these approaches were slightly better than those I had previously done. This motivates me and makes me more confident in using it for a variety of tasks. Naturally, I always do a final check to ensure that everything is correct, cohesive, and without potential issues."\\}

Some developers also mentioned potential negative impacts on their careers, such as creating a higher barrier to entry for novice developers in the market, who could previously be responsible for repetitive and less complex tasks. The following are some comments from the software developers that highlight these aspects:\\

\citatext{\xmark P36: "It would positively save time in most cases, negatively could increase the ruler for a beginner to enter the market, because skills are replaceable by ChatgPT due to low knowledge."\\}

\citatext{\xmark P186: "Maybe the entry point for software engineering will require more skill."\\}

\begin{mybox}{Finding 6}

Developers' views regarding the impact of ChatGPT on their careers exhibit a wide spectrum. While some consider it an avenue for new opportunities, others see it as posing challenges. Specifically, 41\% anticipate the necessity for skill enhancement and adaptation, whereas 16\% foresee minimal impact. Conversely, 38\% hold the belief that ChatGPT will generate job prospects. Additionally, 20\% of respondents intend to integrate it into their work routines, with 36\% already doing so. However, concerns persist regarding heightened entry barriers for inexperienced engineers, as ChatGPT can potentially replace certain skills.
\end{mybox}


\textbf{RQ2.2: What do developers think about the regulation of AI systems like ChatGPT?}

To explore this research question, we asked software developers directly about regulating technologies like ChatGPT. Then, we requested them to explain their opinions, which helped us understand the reasons behind their perspectives.

The responses to the closed question presented various points of view concerning the regulation of artificial intelligence, specifically Large Language Models (LLMs) like ChatGPT. Most of them, represented by 29\% (61/207) of participants, highlighted the need for regulation. 25\%(52/207) of the participants indicated that they had no ultimate opinion on the matter. Approximately 24\% of participants (50/207) considered the possibility of regulation, while 13\% (27/207) were against any form of regulation. Lastly, 8\% (17/207) of the participants expressed concerns about the potential detrimental effects of regulating artificial intelligence.

\begin{figure}[htbp]
    \includegraphics[width=0.6\textwidth]{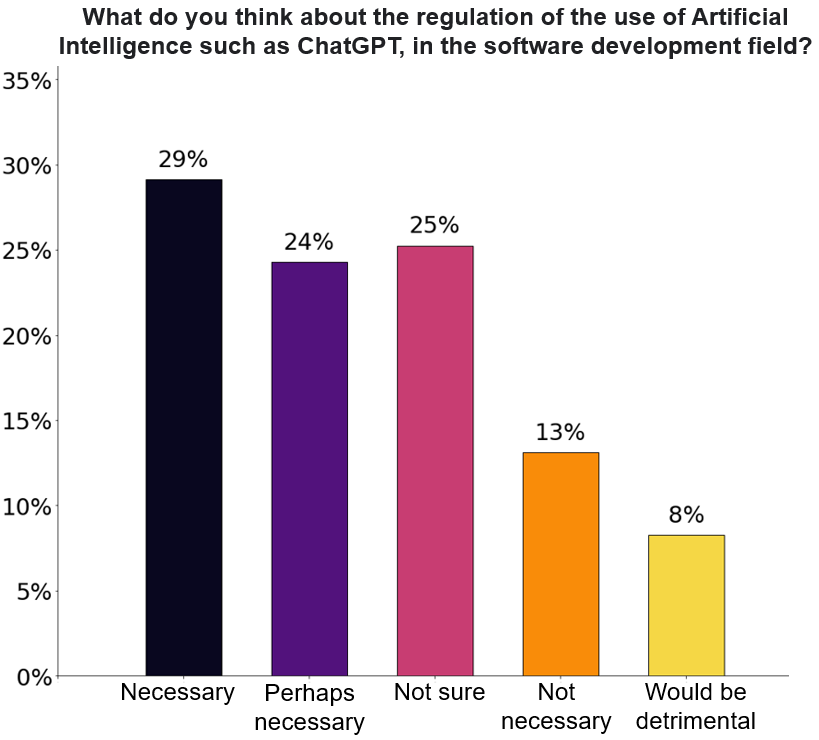}
    \caption{Regulation perception about ChatGPT}
    \label{fig:Regulation perception}
\end{figure}

The open-ended question played a crucial role as a guide, enabling us to comprehend developers' priorities and concerns better. Several recurring themes emerged, highlighting essential aspects to be considered in the debate on the regulation of artificial intelligence.

One of the interesting supportive responses to regulation came from participant P113. He was worried about the technology being misused and the potential negative consequences it could have. \\

\citatext{\xmark P113: "Without regulation, the indiscriminate use of tools like this can bring an avalanche of consequences in the social and ethical scope. These consequences need to be analyzed, and measures need to be taken to ensure that the use of the tool is beneficial for society and industry, rather than something with the potential for harm."\\}

On the other hand, participant P11 made a noteworthy point about the innovative nature of the software industry. He argued that too much regulation could slow technological progress because the industry always changes and develops new ideas. Additionally, he stressed how challenging it is to create regulations that cover every aspect, considering the wide range of uses and constant technological advancements.\\

\citatext{\cmark P11: "Currently, it is believed that regulating the use of artificial intelligence, such as ChatGPT, in the software development field may not be necessary. This is due to the dynamic and innovative nature of the software industry, which could be hindered by overly strict regulations. Additionally, the diversity of use cases and the constant evolution of technology make it challenging to create regulations that encompass all nuances. Instead, many companies and organizations opt for voluntary ethical guidelines to promote responsible practices. However, the debate on AI regulation continues, and opinions may change as technology advances and concerns arise."\\}

In the context of open-ended questions, it was observed that several topics consistently emerged in participants' responses, indicating broad consideration of relevant issues. These topics included security, ethics, privacy, copyright, innovation, regulation, responsibility, society, and the labor market. 
The main points raised by participants who favored regulation and justified their response were: Social impacts 27\% (16/60), Job market 20\% (12/60), Data privacy 20\% (12/60), Copyright 15\% (9/60), Ethics 13\% (8/60), and Security 12\% (7/60). Among those opposed to regulation, virtually all justifications involved the potential delay in innovations and technological development as consequences of regulation.

Regarding ethics, a particularly notable finding was participant P188's concern about accountability. He questioned who would be responsible if the code produced by an AI caused significant harm to a third party or even resulted in death. \\

\citatext{\xmark P188: "Open AI in particular attempts to pass the buck on liability and responsibility for the real world effects created by the output of their software. Another human outside of Open AI makes the decision to send text to Open AI for processing and to do something with the result. Eventually, in some application where there is sufficient lack of oversight in this process and there are monetary damages and/or loss of life, the legal system is going to have to determine which humans and which entities are at fault. I don't believe there are many lawmakers prepared to make effective regulations, and it's hard to tell which ones we're going to need. One day, code written by an LLM is going to be the cause of someone's death and negatively affect the safety, security, and productivity of some nation somewhere. Right now, laws expect a human to be responsible for this task.
If an executive asks an LLM to create a product instead of a human, who will be held responsible?"\\}

Meanwhile, from the perspective of participant P184, the entire responsibility should be attributed to the person deciding to use the code, as he argues that its origin is not crucial but rather the execution.\\

\citatext{\cmark P184: "Doesn't matter if you got a piece of code from ChatGPT, stack overflow, or a team member. It's your responsibility if you decide to use it."
\\}

In discussions about the job market, some responses emphasized the need for job protection. They highlighted developers' concerns about the possibility of being replaced by artificial intelligence, as stated by P136:\\

\citatext{\xmark P136: "I think at least the protection of jobs (which won't happen) is necessary in regulation. I don't think it shouldn't be utilized to its fullest, unfiltered extent."
\\}

Conversely, another group of developers showed no worries about the risk of being replaced. They interact directly with clients and argue that clients themselves may not always have a clear idea of what they want, making direct human interaction crucial in their field, as reported by P99:\\

\citatext{\cmark P99: "ChatgPT is a tool, I see no reason to regulate something that will be used only to help us. The idea that chatgpt would replace developers is at least funny. Product people sometimes don't even know what they are asking, imagine trying to make software themselves without a developer."
\\}

The discussion about jobs also highlighted how programming knowledge can help reduce any problems from using ChatGPT at work. Respondents were worried that new programmers might make code that is hard to understand with ChatGPT's help. Some also mentioned the rules for engineers set by the NSPE (National Society of Professional Engineers). Participant P68 reported:\\

\citatext{\xmark P68: "GPT doesn't fully adhere to all aspects of the NSPE code of ethics for engineers. As such it should be required to regulate the abundant use of GPT for unmonitored automation without final check by a human. Within regulations GPT can be a positive impact but may not be ethically just to allow it to manage large scale projects without oversight."\\}

Finally, some participants strongly oppose new regulations. They believe that current laws already provide enough coverage for the impacts of artificial intelligence. Others argue that we are still in the early stages and that implementing regulations is premature. Some even claim that regulating such technologies is impossible, and there are political reasons related to opposing regulations, as expressed by P121: \\

\citatext{\cmark P121: "Any kind of tool developed with a final purpose must be used by the people to whom this purpose is intended. The state has no right to say which farms can buy a particular machinery or not, just as you do not have the right to determine who you can use, or how you can use, artificial intelligence or not."
\\}

\begin{mybox}{Finding 7}

Most participants who expressed a clear opinion on the topic (53\%) are open to regulating artificial intelligence. Their reasons for supporting regulations include social impacts, copyrights, jobs protection, information filter, security, privacy, safety, source of the information, AI being ``black boxes", ethics, misinformation, and chatGPT hallucinations. Additionally, the other portion of participants expressed opposition to regulation, citing concerns such as hindering innovation, impeding technological progress, dynamic nature of the software industry, and the belief that current laws are adequate to address the impacts of artificial intelligence.
\end{mybox}


\textbf{RQ2.3: What do developers think about the influence of ChatGPT on potential job layoffs in the software market?}

After investigating developers' opinions regarding the potential regulation of ChatGPT and similar technologies, our attention turned to whether they perceive threats to their jobs due to advancements in artificial intelligence. Subsequently, we asked if they believed job layoffs could result from these advances. Additionally, we explored whether they anticipated impacts on the demand for developers and in what way.

For the first analysis, the results were as follows: 8\% (17/207) of participants believe that ChatGPT will not lead to layoffs, 28\% (59/207) anticipate minimal impact, 20\% (42/207) have no formed opinion, 32\% (66/207) believe ChatGPT might cause some layoffs, and 11\% (23/207) think it will certainly result in job losses.

These data reveal that 43\% (89/207) of participants perceive that ChatGPT and similar technologies may directly or indirectly impact future job layoffs, while 36\% (76/207) believe there will be little or no impact.

\begin{figure}[htbp]
    \includegraphics[width=0.6\textwidth]{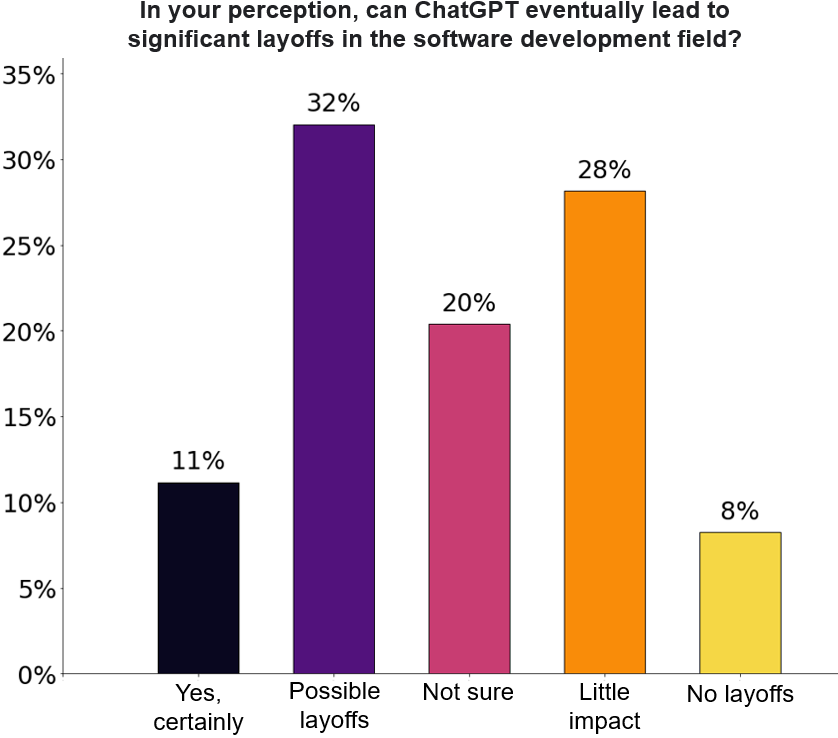}
    \caption{Impact of ChatGPT on Developer Job Layoffs}
    \label{fig:Layoff}
\end{figure}

The next step was to investigate how they thought the demand for programmers would be affected. 3\% (6/207) of participants thought the demand would decrease drastically, while 20\% (42/207) believed there would be some decrease but not as drastic. 54\% (112/207) had no formed opinion or believed that the demand would not be impacted, 19\% (39/207) believed there would be some increase in demand, and 4\% (8/207) believed the demand would increase significantly.

Participant 109 expressed his concern regarding new and future software engineers and the reason leading them to believe that the regulation of artificial intelligence is necessary:\\

\citatext{\xmark P109: "If we don't regulate it, companies will abuse of it and won't invest on new developers for instance, and in the future experienced Engineers or Architects won't have as much experience as we do now. And of course, a lot of Software Engineers won't have places to work."\\}

Participant P137 also argues that the same work can be done by increasingly smaller teams, rendering large developer teams unnecessary for most companies:\\

\citatext{\xmark P137: "Negative: There may be a massive job loss as one person will be able to perform the tasks of many; lower wages may result as the heavy lifting is done by expensive machines; there could be a rapid expansion of mega-corporations as there are fewer limitations on their growth, leading to the cessation of other corporations that can't compete. Positive: There may be more open-source, large-scale, and useful projects; a boom in scientific research may occur as ChatGPT can assist with completing research tasks; there may be short-term positively in software jobs."\\}

\begin{mybox}{Finding 8}
Approximately 43\% of participants anticipate potential job losses due to advancements in artificial intelligence, highlighting concerns about the impact of technologies like ChatGPT on workforce stability. Meanwhile, participants also associate the absence of regulation and the productivity gains from ChatGPT with possible cost reductions for companies, potentially limiting job opportunities for professionals.\\

\end{mybox}


\textbf{RQ2.4: What do developers think about the relationship between ChatGPT and automation in software development?}

\begin{figure}[htbp]
    \includegraphics[width=0.6\textwidth]{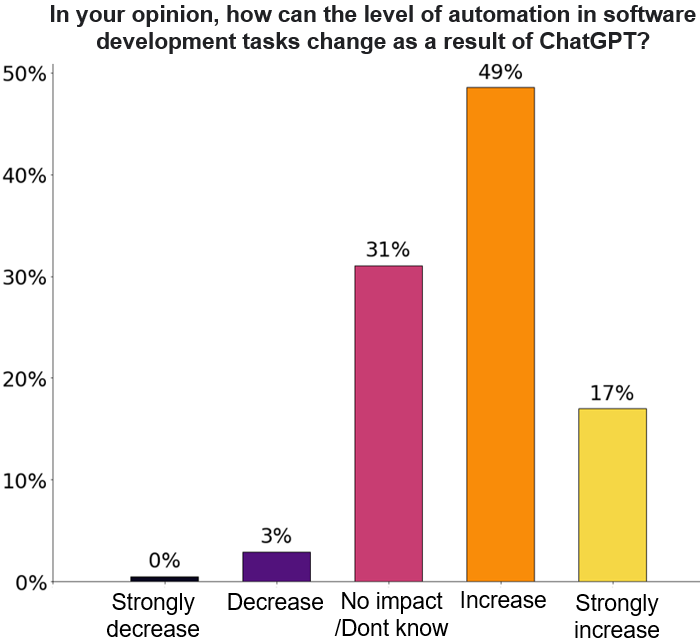}
    \caption{Impact on Automation level due ChatGPT}
    \label{fig:Automation Level}
\end{figure}

In this context, we also aimed to identify opinions on the level of automation in software development regarding the possibilities of using ChatGPT. Our results showed that 49\% (101/207) of participants claimed to believe that there would be an increase in the level of automation in software development due to ChatGPT. 17\% (35/207) of participants said it would increase significantly, while 31\% (64/207) believe there will be no impact. Figure \ref{fig:Automation Level} shows the frequency of each response.

\begin{figure}[htbp]
    \includegraphics[width=0.6\textwidth]{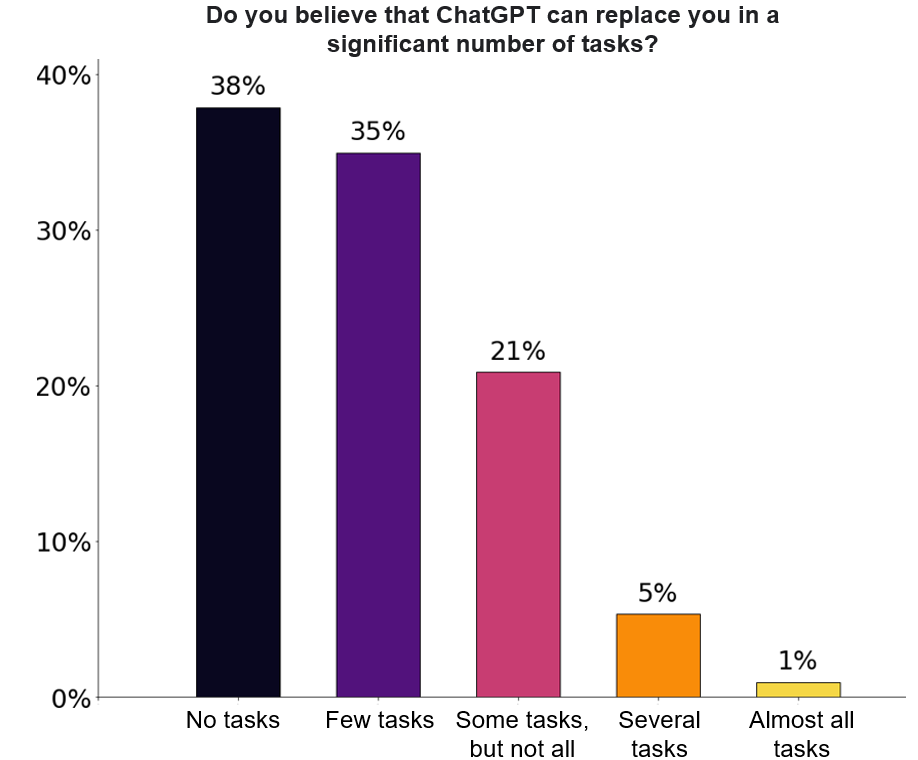}
   \caption{Perceptions on ChatGPT replacement in software development tasks}
    \label{fig:ChatGPT replace on tasks}
\end{figure}

Moving forward with our analysis of automation trends, we investigated how developers feel they could be replaced by ChatGPT regarding the number of tasks. Despite the previous result revealing a strong belief in increased automation in software development, the majority of developers (73\%) demonstrated a belief that they could not be replaced by ChatGPT in any task, or only in a few tasks, as shown in Figure \ref{fig:ChatGPT replace on tasks}.

Analyzing the open-ended responses, we found several comments regarding potential productivity gains and the automation of repetitive tasks as positive consequences. However, participants also generally reported some negative impacts, such as increased review volume and effort in QA stages, as well as the possibility of security risks, dependence, and even dismissals and workforce reduction. An interesting response that highlights it was provided by participant P45:\\

\citatext{\xmark P45: "In my opinion, the adoption of ChatGPT and similar technologies has the potential to increase the level of automation, and here are the consequences: Positive consequences: Increased productivity: Automation of repetitive and low-level tasks, such as simple code generation, can increase the productivity of development teams, allowing developers to focus on more complex and creative tasks. Faster Development: With the ability to automatically generate parts of the code, software projects can be developed faster, which is crucial in a market where launch speed is essential.Negative consequences: Technological dependence: Excessive dependence on automation can make development teams vulnerable to technical failures and compatibility problems when automation tools do not work as expected.Variable Quality: Automatic code generation can lead to variable quality, as tools may not completely understand the nuances of a project. This can result in software that does not fully meet the requirements, or that is difficult to maintain. Security: Poorly implemented automation can create safety vulnerabilities as automated systems may not be as safe as the code carefully revised by humans."\\}

\begin{mybox}{Finding 9}
49\% (101/207) of participants predict more automation in software development with ChatGPT, with 17\% (35/207) expecting a significant increase, while 31\%  (64/207) foresee no change. However, most developers believe they cannot be entirely replaced by ChatGPT. They recognize benefits such as increased productivity and task automation but also express concerns about increased review efforts, security risks, and potential job losses.
\end{mybox}


\textbf{RQ2. Association Analysis.} Investigating the themes related to this research question, we found that the perceived impact of ChatGPT on developers' careers strongly correlates with their willingness to integrate it into their work environments, as evidenced by a significant p-value ($p < 0.001$). We also found that neither experience ($p = 0.832$) nor expertise ($p = 0.606$) exhibited any notable associations with the fear of layoffs. Furthermore, company size demonstrated no significant correlation with job security concerns ($p = 0.226$).

Therefore, we confirmed that none of the evaluated demographic variables have associations with the fear of job layoffs. However, we proceeded with further analysis, given the topic's sensitive nature. Given the recurrence of the topic of layoffs and job protection, we sought to assess whether there is any association between those who fear job layoffs and their opinion on the regulation of artificial intelligence. Using the Fisher method\cite{fisher}, we obtained a $p$-value of 0.003, indicating that there is an association between those who believe in the possibility of ChatGPT impacting job layoffs and those who consider the regulation of artificial intelligence necessary.

We also examined the relationship between company size and their approaches to integrating ChatGPT into software development. Remarkably, larger enterprises, particularly those with over 100,000 employees, were more inclined to impose restrictions on ChatGPT usage compared to smaller ones ($p < 0.001$). This inclination indicates a persistent skepticism among larger corporations regarding ChatGPT's effectiveness, emphasized by their emphasis on software quality and concerns about potential risks associated with its adoption.

Furthermore, our analysis identified intriguing patterns concerning the anticipated heightened automation in software development attributed to ChatGPT. This expectation was more widespread among individuals with limited experience, typically less than a year ($p = 0.029$). On the other hand, those with advanced educational backgrounds were less inclined to anticipate this compared to their peers ($p = 0.03$), suggesting differing perceptions across demographic groups.

Finally, despite the diverse range of responses regarding beliefs about layoffs and task substitution in our analysis, we found no substantial evidence supporting an association between experience level and beliefs regarding the impact of ChatGPT on layoffs ($p = 0.832$) or its potential to replace developers in various tasks ($p = 0.336$). This contrasts sharply with the perspectives of participants who expressed concerns that less experienced developers were threatened or would indeed be supplanted by ChatGPT in the job market. Interestingly, the notion that ChatGPT could lead to layoffs or job displacement does not appear to be limited to a specific segment within the spectrum of software development experience, contrary to the assertions of many participants. Similar observations were made regarding the relationship between regulation and experience ($p = 0.483$), or regulation and ChatGPT usage frequency ($p = 0.704$), with no statistically significant findings in these associations.

\begin{mybox}{Finding 10}
The perception of ChatGPT's positive impact on careers appears to be linked to developers' greater inclination to use it ($p < 0.001$). In relation to job security, we found no statistical significance in associations between concerns about layoffs and factors such as experience ($p = 0.832$), expertise ($p = 0.606$), or company size ($p = 0.226$). However, the belief that ChatGPT causes layoffs was associated with more favorable views on similar AI regulation ($p = 0.003$). Furthermore, individuals with less experience expressed higher expectations for increased automation in software development due to ChatGPT ($p = 0.029$). Additionally, we observed that larger companies tend to impose more restrictions on their developers' use of ChatGPT ($p < 0.001$).
\end{mybox}

\section{Discussion}\label{sec:discussion}
Based on the results in Section~\ref{sec:results}, we discuss implications for researchers and developers.

\subsection{Implications for Researchers}

\textbf{(1) Automation for Complex Programming Tasks:} There is evidence that ChatGPT performs better with simple, repetitive programming tasks than more intricate ones. Researchers are suggested to investigate methods to improve AI models' performance on complex programming tasks.

\textbf{(2) Trade-off Analysis:} The time-saving aspect of ChatGPT usage appears to be substantial but comes with caveats such as heightened review time due to potential inaccuracies. Future research should focus on optimizing the balance between quality assurance and coding efficiency to mitigate negative impacts like hallucination and code errors.

\textbf{(3) Improvement of ChatGPT's Quality in Software Development:} Our empirical finding shows that ChatGPT tends to exhibit for the tasks demanding a deep understanding of code, such as bug-fixing and adherence to security protocol. Furthermore, the model's effectiveness is curtailed by its reliance on outdated training data. Future research is suggested to put more effort into improving ChatGPT's quality of the tasks and mitigating the limitations caused by outdated information.

\textbf{(4) Regulation Development:} The majority support (53.7\%) for AI regulation indicates a pressing need for interdisciplinary research involving legal, ethical, and social implications of AI in software development. Researchers are encouraged to develop guidelines and frameworks that protect intellectual property and user safety.

\subsection{Implications for Developers}

\textbf{(1) Embracing the Automation Trends:} With 50\% expecting an increase in automation, developers should anticipate a shift in responsibilities towards supervising AI systems and integrating them seamlessly into development processes. They should understand that while certain aspects of their jobs may become automated, their expertise remains irreplaceable.

\textbf{(2) Balancing Coding and Reviewing:} Developers should weigh the benefits of reduced coding time against the additional code reviews when using ChatGPT. They are encouraged to prepare themselves for potential changes in employment dynamics by staying informed about the latest trends in AI-assisted development and adapting accordingly. They should also invest in tools and techniques to identify and correct inaccuracies resulting from model limitations.

\textbf{(3) Workflow Adaptation:} Developers should consider incorporating ChatGPT into their workflows for automating repetitive tasks to boost productivity, focusing on code explanation, commenting, and refactoring. However, they must remain vigilant in reviewing AI-generated outputs and avoid over-reliance on AI for critical and security-sensitive tasks.

\textbf{(4) Regulation Engagement:} Developers should actively engage in discussions around AI regulation and its potential consequences, ensuring their voices are heard in shaping the future of the industry. They could collaborate with professional associations and technology advocacy groups to propose guidelines that promote the ethical and secure use of AI while safeguarding job stability and intellectual property rights.

\section{Limitations and Threats to validity}\label{sec:limitations}
In this section, we discuss several threats to validity for our study. 

\textbf{Conclusion Validity} \textendash\
Threats to conclusion validity are concerned with issues relating to the treatment and the study's outcomes, including the choice of sample size and the quality of the material used in the study \cite{Wohlin}.

A pilot study was conducted with two software developers to ensure that the survey instrument was of high quality. Finally, to avoid the threat of drawing false conclusions based on the open questions, we carefully validated our findings with the participants as we performed analysis, asking for clarification when needed.

\textbf{Internal Validity} \textendash\
As a survey-based opinion study without any incentives for participants other than interest in the topic, individuals with more polarized opinions would likely be more inclined to participate. This includes developers who, for some reason, dislike ChatGPT as a tool and do not use it, as well as those who use it all the time. To address this, we included a question about usage frequency within the demographic field to ensure that our research would not be restricted solely to individuals with extreme opinions.

There is also a chance that participants may interpret some questions differently than intended. Therefore, we sought to utilize definitions from other studies whenever possible to avoid subjective interpretations by participants. Additionally, the responses to open-ended questions were individually assessed by two researchers, and after individual evaluation, consensus was reached for coding the participants' responses. This approach aimed to ensure consistency and accuracy in the analysis of participant responses, reducing the influence of potential biases from researchers during analysis.

\textbf{Construct Validity} \textendash\
There are potential threats to construct validity from the lack of a clear definition of a
software developer. Participants generally understood that we meant a member of the development team responsible for coding activities (programming). In addition, we clarified on social networks and online communities whenever there appeared to be confusion.

Another threat to construct validity is related to the potential problem of evaluation apprehension \cite{Wohlin}. It was mitigated by letting the participants know they would remain
anonymous, and by assuring them that all information gathered during the survey would be used solely by the research team and never shared beyond.

\textbf{External Validity} \textendash\ Our survey asked 207 software developers across 31 countries. Though the survey provided important insights, it can be considered a small sample, and the findings may not be generalized to other countries and companies. In addition, as the study focuses on development tasks related to code, we are not capturing how developers perceive the impact of ChatGPT on tasks less related to code, such as requirements elicitation and architecture design. Therefore, our conclusions are restricted solely to specific code-related activities during software development.

\section{Conclusion}\label{sec:conclusion}

Nowadays, Large Language Models (LLMs) have gained significant attention in software engineering, with models like GPT showcasing remarkable capabilities in different tasks ranging from code generation to natural language understanding. However, despite ample quantitative research highlighting the benefits of LLMs in software development, few studies have explored the perceptions of developers who interact with these tools in their work activities.

In this paper, we focus on exploring developers' perceptions about the impact of ChatGPT on software development. By surveying 207 software developers, we have uncovered a nuanced picture of how these advanced AI tools are perceived and integrated into development practices, highlighting the implications for developers and researchers.

As future work, we intend to conduct observation studies in the industry with software developers working on software development projects to obtain more insights about the findings identified in this study.

\bibliographystyle{ACM-Reference-Format}



\end{document}